# Knowledge Organization Systems (KOS) in the Semantic Web: A Multi-Dimensional Review

Marcia Lei Zeng
School of Information, Kent State University, Ohio, USA
mzeng@kent.edu (corresponding author)

Philipp Mayr
GESIS - Leibniz Institute for the Social Sciences, Cologne, Germany
philipp.mayr@gesis.org

## Abstract

Since the *Simple Knowledge Organization System* (SKOS) specification and its *SKOS eXtension for Labels* (SKOS-XL) became formal W3C recommendations in 2009 a significant number of conventional knowledge organization systems (KOS) (including thesauri, classification schemes, name authorities, and lists of codes and terms, produced before the arrival of the ontology-wave) have made their journeys to join the Semantic Web mainstream. This paper uses "LOD KOS" as an umbrella term to refer to all of the value vocabularies and lightweight ontologies within the Semantic Web framework. The paper provides an overview of what the LOD KOS movement has brought to various communities and users. These are not limited to the colonies of the value vocabulary constructors and providers, nor the catalogers and indexers who have a long history of applying the vocabularies to their products. The LOD dataset producers and LOD service providers, the information architects and interface designers, and researchers in sciences and humanities, are also direct beneficiaries of LOD KOS. The paper examines a set of the collected cases (experimental or in real applications) and aims to find the usages of LOD KOS in order to share the practices and ideas among communities and users. Through the viewpoints of a number of different user groups, the functions of LOD KOS are examined from multiple dimensions. This paper focuses on the LOD dataset producers, vocabulary producers, and researchers (as end-users of KOS).

# 1. Introduction

Conventional knowledge organization systems (including thesauri, classification schemes, taxonomies, subject heading systems, name authorities, and lists of codes and terms, produced before the arrival of the ontology-wave) have always been quick adapters of new technologies in their publishing venues and applications. They have had timely appearances in the earliest indexing and abstracting (I&A) databases, online information services, CD-ROMs, Adobe PDF files, HTML websites, and XML databases since the 1950s. Recently they have made their journeys to join the Semantic Web mainstream and turned their products into Linked Open Data (LOD) datasets, along with ontologies that have been developed in the 21st century.

The *Simple Knowledge Organization System* (SKOS) and *SKOS eXtension for Labels* (SKOS-XL) became formal W3C recommendations in 2009, as a separate, lightweight, intuitive language for developing and sharing new knowledge organization systems. SKOS may be used on its own, or in combination with formal knowledge representation languages such as the Web Ontology Language (OWL) (W3C 2009). Eight years later, by the end of 2017, there are over 1000 valid LOD knowledge organization systems (KOS) datasets registered in the DataHub[1], while many LOD KOS services also exist. The KOS products that have adopted the LOD approach using the standardized data model syntax recommended by SKOS and OWL can be found in a variety of domains and formats, from general-purpose to specialized domains, from mono-lingual to multilingual, from classification systems, thesauri, taxonomies to name-authority files, from extracted portions or a complete version of an original vocabulary to the end-products that are made from multiple vocabularies. The release of a LOD KOS product represents a turning point for the producer or provider of a vocabulary; but what are the results?

This paper aims to explore what the LOD KOS movement has brought to various communities and users. These are not limited to the colonies of the KOS constructors and providers, nor the catalogers and indexers who have a long history of applying the vocabularies to their products. Across domains, languages, and places, the LOD datasets creators and LOD service providers, the information architects and interface designers, and researchers in sciences and humanities, are also direct beneficiaries of LOD KOS. After a brief explanation of the term LOD KOS, the features of LOD KOS, and the services providing them (in Section 2 Background), the paper lists the resources used to collect the cases and to cluster user groups based on personas (in Section 3 Methods) which are used to deliver the findings in the main body of the paper. Section 4 "Preliminary Findings" is divided into three sub-sections around three groups: LOD dataset producers; vocabulary producers who are involved in the development and enrichment of KOS, and researchers who are the end-users of KOS. Summaries were given to each of these sub-sections as well as at the end of the paper.

# 2. Background

## 2.1 Explanation of the term "LOD KOS"

Using the terminology of the LOD communities, knowledge organization systems are used as "value vocabularies" (which are distinguished from the "property vocabularies" like metadata element sets). This term refers to its usage in the RDF-based models where the "*resource, property-type, property-value*" triples benefit from a controlled list of allowed values for an element in structured data. A value vocabulary defines resources (such as instances of topics, art styles, or named entities) that are used as values for elements in metadata records. Examples include: thesauri, code lists, term lists, classification schemes, subject heading lists, taxonomies, authority files, digital gazetteers, concept schemes, and other types of knowledge organization systems (Isaac et al. 2011). It is important to remember, however, that a KOS vocabulary is more than just the *source of values* to be used in metadata descriptions: by modeling the underlying semantic structures of domains, KOS act as semantic road maps and make possible a common orientation by indexers and future users, whether human or machine (Tudhope and Koch 2004; for a recent special issue on NKOS see Mayr et al. 2016).

---

[1] https://old.datahub.io/dataset

Another notable term, "light-weight ontologies" refers to those using ontological classes and properties to express the conventional KOS. This is popular among those publishing a thesaurus with an ontology model beside SKOS. Usually they are not considered as "reference ontologies" that have rich and axiomatic theories with the focus on clarifying the intended meanings of terms used in specific domains. In this context, lightweight ontologies are regarded as "application ontologies" which provide a minimal terminological structure to fit the needs of a specific community (Borge et al. 1996; Menzel 2003). Yet the term "ontologies" have been applied to various types of vocabularies, while the approaches such as upper ontologies and hybrid ontologies have been widely applied in generating new KOS.

In this paper, we will use "LOD KOS" as an umbrella term to refer to all of the value vocabularies and lightweight ontologies within the Semantic Web framework. When individual value vocabularies and lightweight ontologies are referenced, the term "vocabulary" or "vocabularies" might be used.

## 2.2 Features of a LOD KOS vocabulary

A LOD KOS vocabulary must follow the principles of Linked Data (Berners-Lee 2006) and must be openly available. The SKOS data model views a knowledge organization system as a *concept scheme* comprising a set of *concepts* (W3C 2009), where each *concept* must be named by a URI (Uniform Resource Identifier) or IRI (Internationalized Resource Identifier). Using a unique identifier to represent an entity or resource is one of the basic solutions for providing machine-processable, disambiguated data. Furthermore, HTTP URIs should be used when releasing a dataset as LOD.

Data of a LOD KOS are expressed as RDF triples and may be encoded using any concrete RDF syntax such as RDF/XML, Turtle, TriG, N-Quads, and JSON-LD, allowing the data to be passed between computer applications in an interoperable way, enabling a KOS to be used in distributed, decentralized metadata applications.

A LOD KOS end-product may be available as a RDF data-dump or accessed through a SPARQL endpoint. Templates for forming SPARQL queries, visualized relationships, on-the-fly mapping/matching services, and other innovative delivery methods may also enrich the presence of LOD KOS on the Web.

## 2.3 LOD KOS vocabulary services

The LOD KOS vocabularies are served by dedicated services. It should be noted that for KOS products, the consistency and synchronization between the original databases and the RDF stores are required. Otherwise, if a KOS's LOD version is not updated when the original data source is updated, then the quality of that product becomes questionable. The following are representatives of widely used, well-maintained service providers (SP). They have developed strategies and technologies to ensure not only the availability but also the interoperability, stability, and scalability of the contents and applications they provide.

For those services which host full content of a KOS vocabulary as well as the management data for each component updated on time, they are also known as vocabulary repositories. The natural languages involved could be monolingual or multi-lingual; the number of KOS vocabularies contained in a repository could range from single one to more than 500. A dedicated portal would provide a unified point of access for KOS vocabularies hosted by a vocabulary service. Some of the services only provide the most current version of a vocabulary, while some maintain all versions. Additional functions might be available in addition to searching, browsing, displaying, and navigating. Some of them also align among vocabularies or provide direct links of data values. The following information of the service providers (SP) is current up to January 1st, 2018.

SP-1. Individual vocabulary's provider.
- E.g., EuroVoc[2], the multilingual thesaurus of the European Union (EU). Terms in EU languages and alignments with eight other KOS are available on website and dump.

SP-2. Individual institution as the provider of all vocabularies produced in the institution.
- E.g., Library of Congress Linked Data Services – Authorities and Vocabularies[3] provides access to all vocabularies promulgated by the Library of Congress including the *LC Subject Headings*, *LC Classification*, and *LC Name Authority File*, plus the many smaller value vocabularies such as various code lists and schemas from the MARC documentation standard, preservation vocabularies, ISO language codes, and other standards.
- E.g., Getty LOD Vocab[4] provides multiple Getty vocabularies, the *Art & Architecture Thesaurus* (AAT), the *Getty Thesaurus of Geographic Names* (TGN), and the *Union List of Artist Names* (ULAN), through both data dump and a SPARQL endpoint, plus a comprehensive list of query templates and documentation. The contents are directly linked to the website of the vocabularies. The *Cultural Objects Name Authority* (CONA) is on its way to becoming LOD.

SP-3. Unified portal for a country's KOS vocabularies produced by multiple units in the country.
- E.g., The Finnish thesaurus and ontology service FINTO[5] enables both the publication and browsing of dozens of vocabularies produced in Finland. In addition, the service offers interfaces for integrating the thesauri and ontologies into other applications and systems.

SP-4. Domain-oriented portal for collected vocabularies produced by multiple units.
- E.g., BioPortal[6] provides a Web portal enabling biomedical researchers to access, review, and integrate disparate ontological resources in all aspects of biomedical investigation and clinical practice (nearly 690). Among the extra features are the mapping among the involved vocabularies, the usage data, and reviews.
- Other examples are: Ontobee[7] (biomedical); Planteome[8] (plants); Ontology Lookup Service (OLS)[9] (biomedical); GFBio terminology service[10] (biological), and Heritage Data[11] (cultural heritage).

SP-5. Middleware that provides tools for end-users to use/reuse published vocabularies.
- E.g., Skosprovider[12] provides an interface that can be included in an application to allow it to talk to different SKOS vocabularies. These vocabularies could be defined locally or accessed remotely through web services, for example, for the Getty vocabularies and the vocabularies published by EH, RCAHMS and RCAHMW at heritagedata.org.

SP-6. Upper ontology that facilitates multiple vocabularies' concept- and entity-mapping.

- E.g., Linked Open Ontology cloud KOKO[13] supports the managing and publishing of a set of interlinked Finnish core vocabularies; enables the users to use multiple ontologies as a single, interoperable, cross-domain representation instead of individual ontologies.
- E.g., Upper Mapping and Binding Exchange Layer (UMBEL[14]) provides an UMBEL vocabulary that is designed for mapping ontologies and external vocabularies (OpenCyc, DBpedia, PROTON, GeoNames, and schema.org), and provides linkages to more than 2 million Wikipedia entities.

Vocabulary registries are different from repositories because they offer information *about* vocabularies (i.e., metadata) instead of the vocabulary contents themselves; they are the fundamental services for locating KOS products. The metadata for vocabularies usually contain both the descriptive contents and the management and provenance information. The registry may provide the data about the reuse of ontological classes and properties among the vocabularies.

SP-7. Registry of KOS.

- E.g., BARTOC[15] (Basel Register of Thesauri, Ontologies & Classifications) currently has over 2740 KOS's metadata in the registry, including active, inactive, or historical vocabularies. 315 of these are available in RDF format. Furthermore, BARTOC includes the metadata of over 80 other registries.

SP-8. Registry of any vocabularies that are published with Semantic Web languages.

- E.g., LOV (Linked Open Vocabularies[16]) currently has over 600 registered vocabularies; all went through certain quality verification. Many of the vocabularies are property vocabularies. In addition to the descriptive metadata about a vocabulary, the usage metadata about properties' reuse among vocabularies, the administrative metadata showing the most recent updates, and the technical metadata regarding the expressivity in terms of RDF, OWL, and RDFS are provided. The details of a vocabulary are exposed through statistics, such as the total number of classes, properties, data types, and instances.

SP-9. Registry of any LOD products, including KOS.

- E.g., DataHub's previous version[17] (as of Sept. 2017) is still the largest registry, with over 11,273 datasets registered. Searching for various KOS types resulted with over 1000, after verification by the authors of the paper.

# 3. Methods

## 3.1 Sources of the study

This study examined cases collected from various sources, including released LOD KOS products, journal articles, conference presentations, workshops and webinars, related tweets, blogs and posts in community-shared spaces. These sources have certain special characteristics worth mentioning here. First, many of the LOD activities are experiments, done outside of the vocabulary creator and indexer circles. Second, in cases where efforts have been initiated by and involve KOS providers, the implementation may take time to be tested, improved and officially added to the workflow. These cases are usually shared within communities and informal groups, especially at the beginning stage of the LOD products life cycle. They are most likely to be publicized through conference presentations, demos, posters, and un-conference sessions, while a smaller number of formal publications appear in journals. Thus, the sources of this research are unconventional and include:

- Sessions of KOS at international conferences
- Research-based journal publications
- Theses and dissertations
- Professional conferences and summits
- NKOS workshops (archived at http://nkos.slis.kent.edu)
- the NKOS bibliography project[18]

---

[13] https://finto.fi/koko/en/
[14] http://umbel.org/
[15] https://bartoc.org/
[16] http://lov.okfn.org/dataset/lov
[17] https://old.datahub.io/dataset

Other sources where cases were exposed to the authors of this paper include:

- LOV[19] on Google+
- Getty Vocab Google Group[20]
- Getty Share[21]
- Social media sources: tweets, blogs, Facebook groups
- Ontolog-Forum[22]
- LODLAM[23] challenges and un-conference-style sessions
- GitHub entries such as OpenSKOS, NatLibFi/Skosmos, JSKOS, and more.

## 3.2 User personas developed for communicating the preliminary findings

In an effort to classify the ideas and outcomes related to LOD KOS reported in the sources listed above, this research first created personas representing typical user groups of LOD KOS in order to build a common understanding of their needs and the goals they wish to achieve. Rather than a top-down approach to collect the definitions of certain user groups, the authors took a bottom-up approach to group the personas that are defined through the project. Although fictional, a persona is a realistic description of a typical or target user of a product, highlighting specific details and important features of a user group. Personas have been widely used in user experience design tasks. They are user models synthesized from real-world observations and are used to incite emphatic thinking when developing a system. It is a process in which data is summarized, clustered, and analyzed to discover themes; the results of which are then used to create outlines or "skeletons" of individual users that can be used for planning, design, and development (Pruitt and Adlin 2010, p.156).

Proto-personas are a modification on traditional personas with the difference that they are not synthesized from data collected from interviews of users. Instead, they originate from brainstorming workshops where company participants try to encapsulate the organization's beliefs (based on their domain expertise and gut feeling) about who is using their product or service and what is motivating them to do so (Gothelf 2012, D'Amore 2016). Proto-personas can be utilized to prevent the design team from viewing themselves as the intended users, and to help guide them create a system suitable for their intended users or user groups (Buley 2013: 132–135, Krøger et al. 2015).

---

[18] https://github.com/PhilippMayr/NKOS-bibliography/
[19] https://plus.google.com/u/0/communities/108509791366293651606
[20] https://groups.google.com/forum/#!forum/gettyvocablod
[21] https://share.getty.edu/display/ITSLODV/Home
[22] https://groups.google.com/forum/#!forum/ontolog-forum
[23] http://lodlam.net/

This study took the approach of proto-personas development based on our literature review and use case studies (using the data sources described in the above section), user behavior observations and brainstorming working group meetings. A number of informal interviews were also conducted. The authors focused on the first tier of persona development defined by Dan Brown (2000): 1) requirements, 2) relationships and 3) humanization). The result is a set of personas encapsulating our understanding of *who* are using the LOD KOS products or services and *what* have motivated them to do so. A persona group, e.g., Vocabulary Producer (VP), contains multiple personas such as VP1, VP2, VP3, etc., they are highlighting different roles of the VP group, might take in one or more projects, or in the same project over time. Among the five groups, the first three will be used in this paper:

- LOD Dataset Producer (DP) group
- Vocabulary Producer (VP) group
- Researcher (RS) group (as end-users)
- Website/Tool Developer (WD) group
- KOS Service Provider (SP) group

The formation of personas follows common practice in that they are very brief, typically bulleted lists of distinguishing data ranges for each subcategory of a user (Pruitt and Adlin 2010, p.184). The resulting personas are intentionally simple and depict: (a) who the group is, including the name and identity key of fictional characters; (b) what are the sources of characters; (c) which tasks they usually have; (d) what are the contents they deal with; (e) where and how they interact with the KOS vocabularies; and (f) what are the goals (see Appendix A for one example of the Vocabulary Producer (VP) persona document). We consider these "skeletons" of the personas to be living documents that support this particular research which uses unconventional data resources, while allowing the profiles to be further refined, split into narrower personas, and encompass more personas as new details are discovered at any time. They are used to provide a central point to enable us to communicate the preliminary findings and to share the cases around LOD KOS.

## 4. Preliminary Findings

Through the viewpoints of different personas designed in this study, the functional changes (or other changes if found) of KOS after they were released as LOD are examined from multiple dimensions. The following sections are organized around personas representing typical users of LOD KOS. Even though some specific cases are used as examples, the attention is on summarizing the general issues and benchmarks identified by this study. Best practices acknowledged by communities as well as experimental approaches are presented, together with the possible challenges and hurdles.

### 4.1 For LOD Dataset Producers (DPs), LOD KOS vocabularies enable their data to become 4-star and 5-star Linked Open Data

In this part, the preliminary findings are presented for LOD dataset producers (DPs) facing different levels of situations when producing LOD products: 1) creating LOD datasets from scratch and dealing with data that have no controlled values for the named entities and topics; 2) reaching out to the datasets that may have or have not been using community standard vocabularies in their structured data; and 3) turning the existing datasets that have been using value vocabularies into 4-star and 5-star LOD.

Before looking into this section, it is necessary to revisit Tim Berners-Lee's 5-star Open Data Scheme for LOD data (Berners-Lee 2006).

| | |
|---|---|
| ★ | Available on the web (whatever format) but with an open license, to be Open Data |
| ★★ | Available as machine-readable structured data (e.g. excel instead of image scan of a table) |
| ★★★ | as (2) plus non-proprietary format (e.g. CSV instead of excel) |
| ★★★★ | All the above plus, Use open standards from W3C (RDF and SPARQL) to identify things, so that people can point at your stuff |
| ★★★★★ | All the above, plus, Link your data to other people's data to provide context |

Among the datasets found in the Datahub, which mostly qualify to be the 4-star, only about 10% were included in the LOD Cloud 2017-02 version as the recognizable 5-star datasets. One of the reasons is that "The dataset is not interlinked with other datasets" (http://lod-cloud.net/, 2017). The LOD KOS vocabularies are primary sources which enable datasets to become 4-star and 5-star Linked Open Data. This benefit has become the most widely acknowledged by the LOD dataset producers.

The LOD dataset producers are dedicated to exploit existing data and deliver structured data in the RDF format. They might be dealing with already structured data such as bibliographic records, museum documentation files, clinical trial databases, etc. More often, they would make structured data out of unstructured raw data such as oral history transcripts. In order to break the silos and connect with the rich information outside of their silo boundaries, many of them took the Linked Data approach and opened up. The linking points are primarily the concepts and named entities, i.e., the identifiable things including people, organizations, places, events, objects, concepts, and virtually anything that can be represented in structured data (see a recent example in Binding and Tudhope, 2016). In the RDF triples (*subject-predicate-object*), they occupy the positions of *subject* and *object*.

Nevertheless, for a dataset to become real LOD, identified entities need to be named with URIs. This is usually the first hurdle to overcome. Thus, using LOD KOS has become a best practice and popular strategy for the LOD dataset producers. Depending on the situation (see Figure 1), the usage of LOD KOS might involve multiple choices and steps.

Figure 1. The options and actions related with KOS in the LOD dataset production

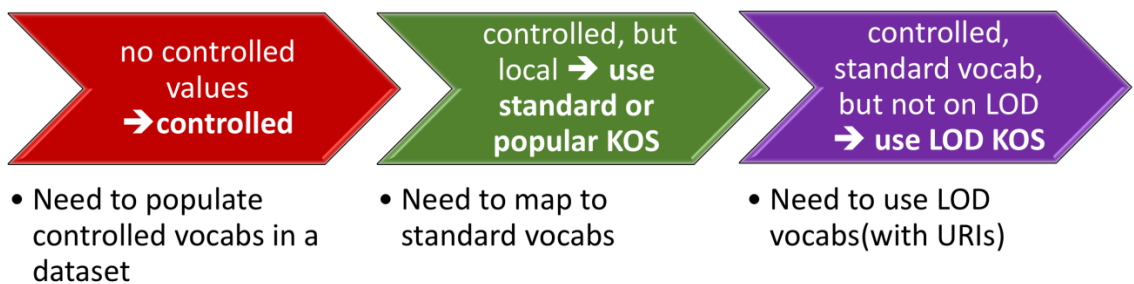

- Need to populate controlled vocabs in a dataset
- Need to map to standard vocabs
- Need to use LOD vocabs(with URIs)

**DP-1. Dealing with semi-structured and unstructured data that have <u>no controlled values</u> for the named entities and topics in order to create LOD datasets from scratch**

Dataset producer DP-1 is dealing with semi-structured and unstructured data that have <u>no controlled values</u> for the named entities and concepts and wants to create a LOD dataset from scratch. The examples of such kinds of data include: the digitized materials (textual or non-textual) hosted in silos; archival finding aids; oral history transcripts; merged local files and others. Technologies exist to help mining data and extracting the entities. However, there are many possible issues to be faced from the beginning. Examples among those involving place related entities are:

- Place names change through time (e.g., "Bombay" and "Mumbai");
- Alternate names exist (e.g., "New York City," "NYC," and multilingual labels);
- Same name is associated with multiple locations (e.g., various "St. Petersburg" in the world);
- Unidentifiable places (e.g., places referred in a creative work but has not been identified);
- Unnamed places;
- Cartographic versus geographical placement; and
- Feature typing/categorizing results are incorrect or inconsistent.

In the effort to identify and control the named entities and concepts from these semi-structured and unstructured data, and advance from digitization to datafication, these major benchmarks are to be reached before becoming 4-star data:

1. Identify the entities;
2. Put the entities into structured data;
3. Clean up the newly structured data, with local control;
4. Encode the entities with standardized KOS vocabularies (as strings);
5. Obtain URIs for entities provided by the LOD KOS datasets; and
6. Use http URIs for names of any entities.

The last three are related to LOD KOS use, in order to have high quality and trustable linkages in the RDF triples.

A well-known pioneer case is Linked Jazz[24], which concentrates on a special collection of Jazz musicians' interviews (Pattuelli 2012; Pattuelli et al. 2015). Based on the data about individual musicians, the team made connections between people. Step 1 was to get the names from the transcripts and establish name authority file with URIs. A natural language processing tool pulled entities from the transcripts of interviews with jazz musicians that mention a relationship with another jazz musician. After the process of controlling synonyms and eliminating ambiguity, the musician names were mapped to name authorities of Virtual International Authority File (VIAF), LC name authority, and DBpedia, and the data about each person was obtained. If a name was not in the name authority, the team established the authority record for the person. Step 2 was to find the names in all relevant documents in the collection based on the established name authority file. Step 3 was to describe the relationships using a relationship ontology the team developed. Finally, a visualization tool was used to present a unique interactive interface.

**DP-2. Reaching out to the datasets that <u>may have or have not been using community standard vocabularies</u> in their structured data**

An effort that needs to integrate distributed data sources from outside institutions most likely will face the issues of standardization or unification on data models and value vocabularies. In this situation, the dataset producer DP-2 intends to reach out to the datasets that may have or may have not been using community standard vocabularies in their structured data. One of the key tasks involves the conversion of existing KOS into LOD before applying them as standard value vocabularies in all datasets to be integrated.

An example of such a situation was reported by the project of Archives of France (Sibille-de Grimoüard 2014). The *Thesaurus W. Standardized vocabularies for describing and indexing local administration records*, developed in 1987, has been used by the French archival agencies to index descriptions of modern records created by local public services. The thesaurus and three controlled lists of terms were available as Excel sheets and PDF files on the Internet till 2008. The ability to interact with the applications used by local archival institutions would need machine-readable and machine-processable KOS. The following needs were identified when the

---

[24] http://linkedjazz.org/

project initiated the LOD activity:

- Represent the thesaurus in a machine-understandable way for automating machine-assisted indexing processes;
- Facilitate its integration into retrieval tools;
- Ensure the consistency of indexing even though the thesaurus evolves;
- Facilitate the process of updating and maintaining the thesaurus (evaluating the requests for changes from users, updating terms and relationships, amending terms, customizing the display of terms, etc.);
- Express all the concepts already represented in the thesaurus (concepts and terms, relationships between these concepts, annotations, etc.); and
- Use standards and models related to thesauri and controlled vocabularies for interoperability purposes (Sibille-de Grimoüard 2014).

This is a very well summarized list of tasks and reflects the needed benchmarks of many projects that may deal with local and distributed sources of data. Even though the thesaurus was not considered fully compliant with ISO-25964 (2011, 2013) as a "thesaurus," the SKOSified KOS enabled the dataset producers to reach the stated goals. This project was also an opportunity to align data with other LOD KOS and resources (e.g., RAMEAU and DBpedia) and to implement a solution for persistent identifiers of concepts of the thesaurus. Among the advantages for users were that the shared use of common vocabularies creates interoperability without any additional developments. For instance, as the thesaurus for indexing local archives provides links to RAMEAU, it would be possible to link an archival resource and a library book through these two thesauri and the links they share (Sibille-de Grimoüard 2014). A similar example of converting a thesaurus into SKOS in the Social Sciences was reported by Zapilko et al. (2013).

**DP-3. Having datasets that <u>have been using value vocabularies</u> in structured data, turning them into 4 star and 5-star LOD**

Dataset producer DP-3's objective is to turn the existing datasets into 4-star and 5-star LOD. These datasets have been using (born-with or mapped-to) value vocabularies in their structured data. Examples of such data include the national bibliographies, catalogs, special collection portals, metadata repositories, and many theme-based LOD products made in projects. A new dataset's resource may be maintained by different information systems based on traditional relational data models. In such a situation, a dataset usually has controlled the named entities and topics with KOS vocabularies. The following benchmarks are expected before becoming 4-star and 5-star data:

1. Use standardized protocols for metadata structure;
2. Enrich the original metadata, especially for those semi-structured and non-controlled fields;
3. Control the value spaces for all entities;
4. Encode the entities with standardized KOS vocabularies (as strings);
5. Use URIs for names of entities; and
6. Use http URIs for names of any entities.

The fifth and sixth benchmarks require that the KOS vocabularies being used are LOD datasets themselves. Fortunately, most of the standardized KOS vocabularies have become LOD KOS. Otherwise the last two benchmarks might not be reachable. However, there are many possible issues for each of the datasets currently at the 3-star level, as summarized below:

- If it has used local controlled vocabularies, the *terms* used or the form representing the concepts and named entities may be different from standardized controlled vocabularies.
- If it has used pre-LOD vocabulary, there might be no *URIs/IRIs* yet. How to obtain the URIs/IRIs to replace the strings of a named entity or concept?

- If a decision of *mapping* is made, which vocabulary and how many vocabularies will be involved, since in a subject domain and a community there could be more than one standard vocabulary.
- If it needs to map the local controlled lists to a standardized LOD KOS (e.g., LCSH, EUROVOC, etc.), human resources and *quality control* are most critical and could be challenging.
- For a dataset formed through *aggregation*, in addition to the above issues, synonyms and homographs occur in the data provided by different sources. Heavy disambiguation and semantic conflict controls are needed.

There are no black-and-white answers to these questions. Many dataset producers developed their own successful products, such as the national bibliographical databases, OCLC's WorldCat, and many others that used various KOS to become 5-star LOD data (Figure 2).

Figure 2. The 5-star LOD Cloud indicates the essential role of LOD KOS vocabularies.
Source: Annotated by the author on the LOD CLOUD 2014-08-30 image
http://lod-cloud.net/versions/2014-08-30/lod-cloud_colored.png

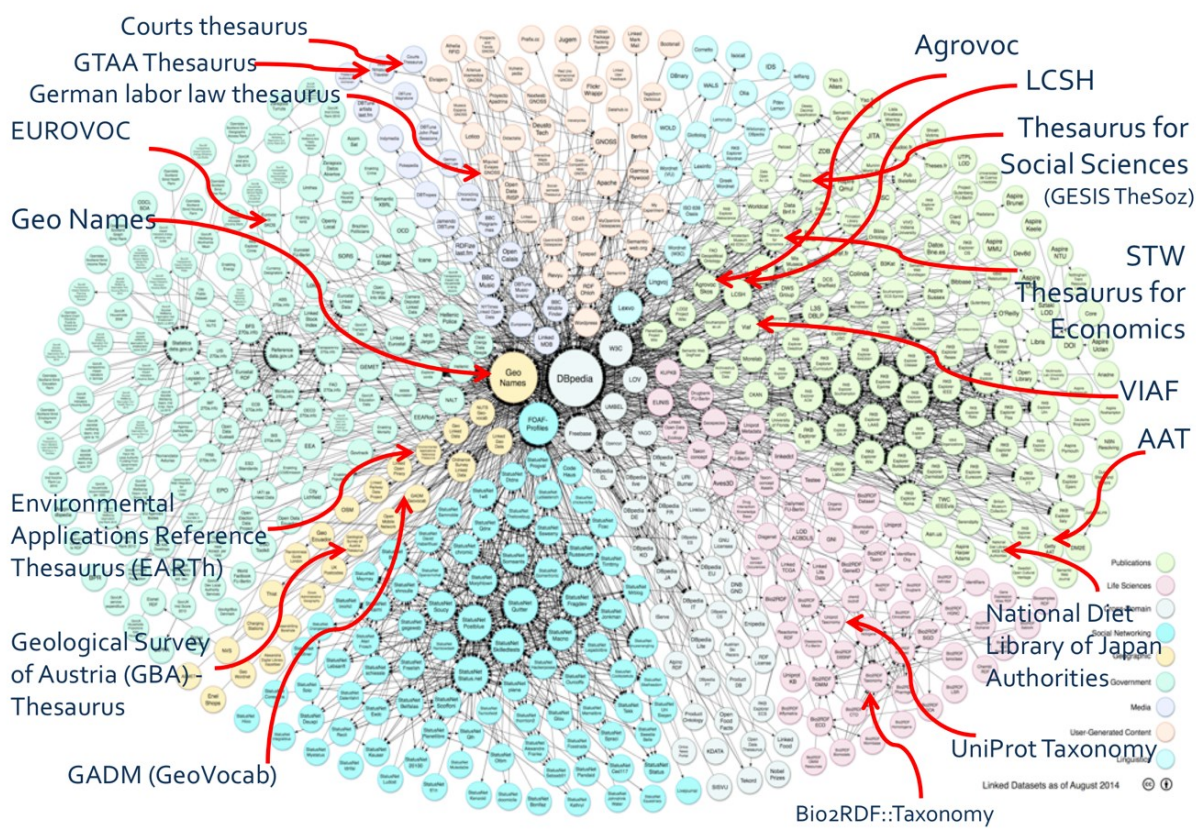

**Summary of usages and practices (Dataset Producers)**

It is clear that LOD KOS vocabularies, as the source of http URIs/IRIs for named entities and concepts used in data-transformation, enable the dataset producers to make 4-star or 5-star Open Data. In the bibliographic universe, they help the conversion from *Library Entities* to the *Web of Data* (Wallis 2014). The possibilities for linkage of high quality structured data become limitless and show the impact in the increased availability of information.

LOD KOS vocabularies empower the owners of data to convert and publish their data under the LOD principles, with high quality and trustworthy linkages in RDF triples. LOD KOS can be used to transform anyone's database into LOD datasets, even reaching 4- and 5- stars; to create machine-understandable and machine-processable data for any users, machine or human. We all understand that the creation of a KOS vocabulary involves tremendous intellectual efforts and human resources, thus, the openly available, well-established, and constantly-maintained vocabularies are invaluable engines for the LOD datasets.

## 4.2 For Vocabulary Producers (VPs) who are involved in the development and enrichment of KOS, LOD approaches lead to unconventional processes and results

The goals of vocabulary producers (VPs) include creating needed value vocabularies for their datasets, while also aiming at sharing the products with communities. The tasks of development and enrichment of new or existing KOS vocabularies are closely related to what LOD dataset producers usually encounter, as presented in the previous section of this paper. The value vocabulary producers to be discussed in this sub-section are considered to be different from the usual vocabulary providers such as those working for a thesaurus or classification system as editors. The following cases are presented based on five objectives: 1) creating new value vocabularies for particular project's products by extracting the components from a comprehensive KOS vocabulary; 2) creating a unified scheme for a domain based on multiple KOS vocabularies; 3) creating a heterogeneous meta-vocabulary; 4) enriching the KOS-at-hand and connecting to real things; and 5) enhancing semantic consistency of data through shared, unconventional mashup Knowledge Organization activities. Despite that the presented cases resemble the approaches used in the KOS community for a long time, new methods, functions and results are observed in current approaches.

**VP-1. Creating new value vocabularies for particular project's products by extracting the components from a comprehensive KOS vocabulary**

When a particular project does not need to apply a full standard thesaurus, or when one existing thesaurus is not enough for the project's domain coverage, extracting components from standardized KOS vocabularies can be a relevant strategy. For example, the Government of Canada's Department of Canadian Heritage – Canadian Heritage Information Network (CHIN)'s *CHIN Guide to Museum Standards* (last updated 2016-09) provides a list of vocabularies of the terminologies for object naming; materials and techniques; disciplines; and styles, periods, and cultures. Each of these terminologies can be a portion of the *Art and Architecture Thesaurus* (AAT). A new vocabulary can be considered as a microthesaurus, which is a designated subset of a thesaurus that is capable of functioning as a complete thesaurus (ISO25964-2:2013).

Vocabulary producer VP-1 is committed to creating new value vocabularies for particular project's products by extracting the components from a comprehensive KOS vocabulary. Whether VP-1 is extracting a whole facet from AAT (e.g., *Object Facet*), a sub-category under a guided term (e.g., < *Object genres by function*>), or a specific group [(e.g., *ceremonial objects* or *vessels (containers)*], the creation of a microthesaurus, with all the components and their RDF triples and URIs, can be obtained by querying AAT through a SPARQL query endpoint, using a template already provided to trace data of "Descendants of a Given Parent." The dataset can be obtained in about two seconds after a query is submitted (Garcia et al. 2017; Zeng 2017) (see Figure 3 and 4).

Figure 3. Using a template provided to trace data of "Descendants of a Given Parent"
for "<costume by function>" (AAT concept ID 300212133).
Source: http://vocab.getty.edu/queries#Descendants_of_a_Given_Parent

Figure 4. Querying for "<costume by function>" (AAT concept ID 300212133), receiving and downloading the datasets to make a microthesaurus.

Based on the sources of the study, especially at the Q&A portion of the conference sessions and community shared spaces, some vocabulary producers expressed concerns regarding limited knowledge of the new semantic technologies such as: #1, dealing with SPARQL queries and using the endpoints; and #2, handling the vocabularies in RDF formats. Aiming at the #1 concern, some middleware (e.g., Skosprovider) provide tools for end-users to use/reuse published vocabularies. Other LOD KOS service providers (SPs) mentioned in Section 2.3 of this paper also provide various tools for constructing, reusing, and enhancing vocabularies. The cases to be discussed in Section 4.3 for researcher (RS) end users might be the best solutions to help these vocabulary producers. The #2 concern regarding handling the vocabularies in RDF formats is common, since a VP would need to organize and edit the selected concepts and terms before finalizing a set of entries to form a needed vocabulary. To solve this issue, the LOD KOS services usually offer multiple downloading formats to be selected by an end-user. One of the commonly used non-proprietary formats is CSV, a comma-separated values (CSV) format. A CSV file stores tabular data (numbers and text) in plain text, which allows a user to open the file from a spreadsheet to work on it directly. CSV is also the preferred form for visualization tools such as Google Fusion Tables and Gephi.

**VP-2. Creating a unified scheme for a domain based on multiple KOS vocabularies**

The Semantic Web encourages the sharing and reuse of data, including the components of KOS vocabularies. The above query example for VP-1 is applicable when obtaining any components of a LOD KOS vocabulary. It is also practical and common to form a new vocabulary based on more than one source, as the vocabulary producer VP-2 is engaged. The following cases demonstrate innovative approaches and results.

*Thesaurus of Plant characteristics* (TOP[25]), which complies with Semantic Web principles, is committed to the harmonization and formalization of concepts for plant characteristics widely used in ecology (Figure 5). It builds on previous initiatives and vocabularies for several aspects, including its model, entities and qualities, and concept definitions (Garnier et al. 2017). TOP provides names, definitions, formal units and synonyms for more than 700 plant characteristics.

Figure 5. An entry of TOP presenting the definition with multiple sources of the concepts, coded with the URI from the original namespaces (e.g., *PO:*, *EFO:*, *Mayr:*).
Source: http://www.top-thesaurus.org/annotationInfo?viz=1&trait=Frost%20tolerance

Motivated by the notion that open data needs common semantics for linking diverse information, the *Global Agricultural Concept Scheme* (GACS) project of Agrisemantics aims to create a shared concept scheme by integrating existing standard vocabularies in agriculture and environment (Baker et al. 2016a). Agrisemantics is an emerging community network of semantic assets relevant to agriculture and food security. GACS functions as a multilingual thesaurus hub that includes interoperable concepts related to agriculture from several large KOS: *AGROVOC* of the Food and Agriculture Organization of the UN, the *CAB Thesaurus* by CAB International of UK, and the U.S. *National Agricultural Library (NAL) Thesaurus*, all maintained by different institutions. GACS would facilitate search across databases, thereby improving the semantic reach of their databases by supporting queries that freely draw on terms from any mapped thesaurus, and achieving economies of scale from joint maintenance. The latest GACS beta version provides mappings for 15,000 concepts and over 350,000 terms in 28 languages as of its May 2016 release (Baker et al. 2016a). The case reveals unique processes and designs: 1) The mappings focused on three sets of frequently used concepts (10,000) from each of the three partners. 2) Mappings were automatically extracted and then manually evaluated by experts through discussions and manually corrected. 3) A classification scheme that was developed jointly in the 1990s was revised to tag concepts by thematic group (chemical, geographical, organisms, products, or topics). 4) Alongside generic thesaurus relations to broader, narrower, and related concepts, organisms will be related to relevant products (Baker et al. 2016b).

Around the world, activities of creating a unified scheme for a domain, focusing on generating multilingual labels by using SKOS-XL, have been proved to be successful, as reported by many other cases.

---

[25] http://www.top-thesaurus.org/

**VP-3. Creating a heterogeneous meta-vocabulary**

Vocabulary producer VP-3's task is similar to VP-2's task discussed above in generating a product based on multiple existing vocabularies. However, the situation involves creating a heterogeneous meta-vocabulary that supports the representation of changes and differing opinions of certain concepts. The case used here is a taxonomic meta-ontology *TaxMeOn,* built by Tuominen, Laurenne, and Hyvönen (2011). *TaxMeOn*[26] is an ontology schema for biological names, containing 12 ontological classes with 49 subclasses. The datasets utilized in the study consist of 20 published species checklists that cover mainly northern European mammals, birds and several groups of insects, resulting in about 78,000 taxon names. The difference between *TaxMeOn* and the cases shared with VP-2 is that the representation of the dataset encompasses these contents: 1) the different conceptions of a taxon, 2) the temporal order of the changes, and 3) the references to scientific publications whose results justify these changes. The rationale is that the positions of species and the nomenclature in scientific taxonomies involve a lot of changes, which directly impacts the access to the publications and data associated with them in different time periods.

The direct application of the taxon meta-ontology model that allows multilingual, different opinions for the biological taxonomy concept and nomenclature in a unified view can be beneficial to the researchers of biology. The detailed data can be further linked to other datasets with less taxonomic information, such as species checklists, and provide users with more precise information. The data model enables managing heterogeneous biological name collections and is not tied to a single database system (Tuominen et al. 2011). More importantly, this modeling method and the model itself can be extended in a flexible way and integrated with other data sources.

**VP-4. Enriching the KOS-at-hand and connecting to real things**

Vocabulary producer VP-4 has a SKOSified thesaurus at hand and is investigating how and when to link a concept in the thesaurus to the URIs provided by name authorities and Wikipedia so as to fully benefit from LOD and enrich an existing KOS-at-hand. Another question is how to take advantages of such processes to allow any organization to improve and expand the data with other relevant sources the organization does not own. For years, there have been discussions about whether name authorities should be maintained separately from concept-based subject heading lists, thesauri, and classification schemes that also contain named entities. In the Linked Data movement, there have been confusing and incorrect applications of skos:exactMatch and owl:sameAs to align *a real thing* (e.g., a person, institution, or place) to the concepts, names, or photos that *represent the thing*.

FAST[27] (Faceted Application of Subject Terminology), a joint vocabulary effort of OCLC and Library of Congress, based on LCSH, reported using *foaf:focus* to allow FAST's controlled terms (representing instances of skos:Concept) to be connected to URIs that identify real-world entities specified at VIAF, GeoNames, and DBpedia. With the correct coding of properties, machines can understand (reason) that a FAST's controlled term is related to a real-world entity and allows humans to gather more information about the entity that is being described (O'Neill and Mixter 2013). As Schema.org[28] grows, classes and properties defined by it are also being applied to FAST. The enrichment allows FAST terms to take advantage of all of the various string values included in VIAF (containing dozens multilingual name authorities) without having to manually include the values in the RDF triples for the specific term in FAST. The GeoNames data is used to power MapFAST, which is a Google Maps mash-up. The DBpedia identifiers allow FAST terms to include detailed information that is usually excluded in authority records (O'Neill and Mixter 2013).

---

[26] http://onki.fi/onkiskos/cerambycids/
[27] http://www.oclc.org/research/activities/fast.html
[28] http://schema.org/

Bensmann, Zapilko and Mayr (2017) reported another large-scale interlinking project in Swissbib[29], a provider for bibliographic data in Switzerland. Data available in Marc21 XML were extracted from the Swissbib system and transformed into an RDF/XML representation. From approximately 21 million monolithic records, the author information was extracted and interlinked with authority files from the VIAF and DBpedia. A main obstacle was the amount of data and the necessity of day-to-day (partial) updates. As a result, the team has developed procedures for extracting and shaping the data into a more suitable form, e.g., data are reduced to the necessary properties and blocked (see Figure 6). The approach could establish 30,773 links to DBpedia and 20,714 links to VIAF and both link sets show high precision values and could be generated in reasonable expenditures of time, according to the authors.

Figure 6. Data flow diagram of the interlinking procedure in the Swissbib project
Source: Bensmann et al. (2017), p.8 Figure 4.

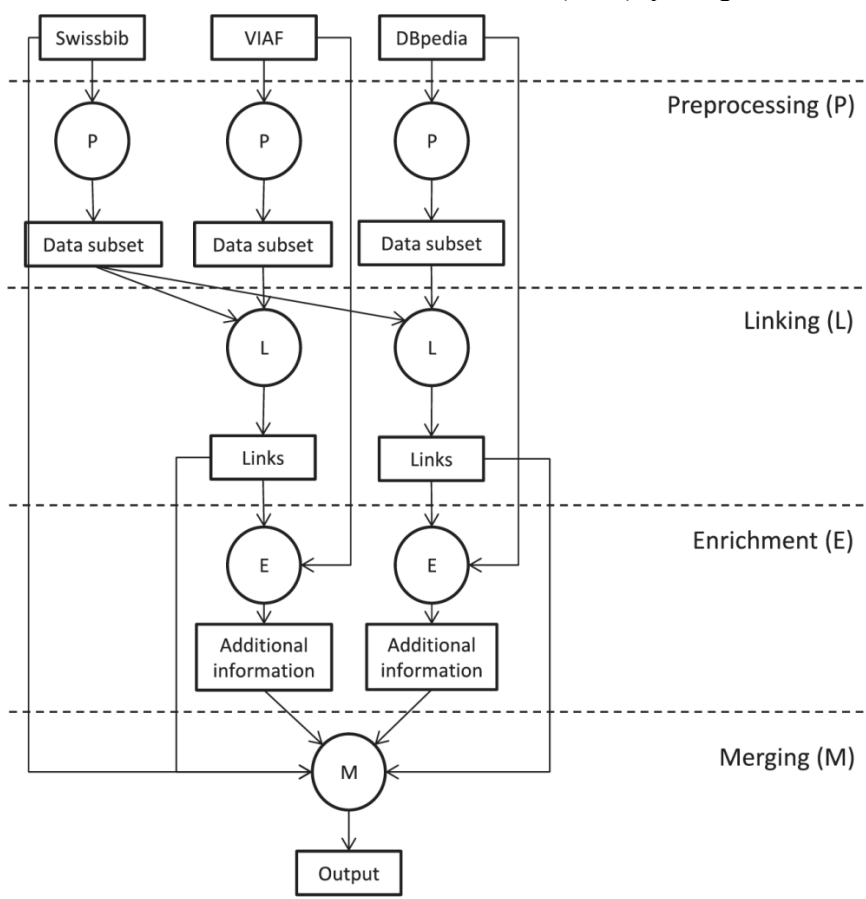

**VP-5. Enhancing semantic consistency of data through shared, unconventional mashup Knowledge Organization activities**

Vocabulary producer VP-5 is involved in the new efforts to enhance semantic consistency and interoperability through shared data which have already shown great potential for dataset producers and KOS vocabulary producers. In Web development, the term "mashup" denotes a combination of data or functionality from two or more external sources to create a new service. "Mashup culture" puts a cultural dimension into the foreground, as these developments permeate through almost all cultural techniques and practices on a global scale (Sonvilla-Weiss 2011).

The most obvious cases are the added authority identifiers and categories in *Wikipedia* entries. The "Authority Control" section has been added to many Wikipedia pages providing the identifiers from name authorities such as WorldCat Identities, VIAF, ISNI, ULAN, etc.

The more systematic activities can be found around *Wikidata* (https://www.wikidata.org), which functions as the authority files of named entities, but increasingly more abstract concepts have been added by volunteers. For instance, the Wikidata Visual Arts project, which intends to provide a knowledge base of reusable multilingual facts that can be used in Wikipedia and any other sites, provides the *Visual Arts Item Structures* as the guideline and classification for describing information related to visual arts. Each entry for an entity has a URI and the classes align with AAT (mostly the <Objects> facet), using AAT URIs as identifiers. Identifiers from other KOS and collections are also found for various concepts[30]. Similar projects can be found across Wikidata, Wikimedia, and other shared efforts.

Although the overall quality, coverage, and mapping accuracy have not been systematically measured or proved, and the sustainability and consistency applied to each concept and named entity are not standardized, these unconventional, shared knowledge organization activities certainly provide a good reference source and quick access to LOD KOS products, filling up the gaps between currently existing KOS coverages and real world's needs. The mash-up culture, a symptom of a wide paradigm shift in our engagement with information, seemed to be perfect for the data-driven cultural techniques and practices of knowledge organization (Voss 2013, Sonvilla-Weiss 2011, Bensmann et al. 2017).

**Summary of usages and practices (Vocabulary Producers)**

The cases presented so far for the Vocabulary Producers (VPs) seemed to resemble KOS methods developed prior to the 21st century. From conceptual and structural points of view, the newly generated vocabularies, derived from the existing ones, took similar approaches such as making microthesauri and satellite vocabularies, creating a super structure, direct mapping or employing a switching system, crowd-sourcing, post-control, etc.

The new functions and differences observed in current approaches are the results of applying LOD principles. Each *thing* included in all the new products is named with a URI, and has a domain name prefix that directly indicates its origin, thus, maintaining the original semantics and linguistic decisions while being reusable.

The cases also benefit from semantic technologies and the available open tools. For example, the new microthesauri or satellite vocabularies can be generated through modifiable SPARQL queries and obtain datasets in a minute. The variety of downloadable formats available allows for easy integration with other data and visualization using open tools.

For vocabulary producers, the LOD KOS vocabularies are the resources for creating, maintaining, enriching, extending, and translating a value vocabulary that complies with LOD principles. The data-driven, shared editing and publishing workflow also facilitates the capture of administrative, provenance, and uses metadata for the whole vocabulary and its components. With an increasing number of KOS published in standardized, machine-understandable formats, it becomes necessary for organizations to improve and expand the KOS data that they already have by using other relevant sources.

---

[30] Refer to https://www.wikidata.org/wiki/Wikidata:WikiProject_Visual_arts/Item_structure

The most important achievement is the reusability of any of these new vocabularies in LOD or non-LOD databases. As the Agrisemantics project team determined (Baker et al. 2016b):

- Open-access semantics are easy to re-use;
- Mapping the semantics promotes cooperation and reduces duplication; and
- Coherent semantics benefit research, innovation systems, and value chains.

## 4.3 For Researchers (RS) who are end-users of KOS, LOD KOS products can become knowledge bases and provide semantic-rich discoveries

It is very common that real end-users (i.e., those other than the creators and publishers of KOS products) may not be familiar with KOS and may not be tech-savvy. The question of how to attract users and extend beneficiaries further than the dataset producer (DP) and vocabulary producer (VP) groups is a major challenge for the LOD KOS vocabulary service providers. Especially they seek to demonstrate the societal value of their efforts of converting KOS into LOD format and providing services such as free data dumps and SPARQL endpoints (which may add extra costs). For this reason, they need users and supporters from all disciplines.

What is more, the scalability of LOD approaches in relation to KOS must be addressed. The data dumps (which are the most popular for LOD KOS) and SPARQL endpoints seem not to be applicable for end-users whose jobs are not related to semantic technologies. Technologically, in addition to the access issues related to finding, browsing, and navigating within or across KOS vocabularies, the challenge arises as to how the LOD KOS can be used as more than traditional "controlled vocabularies" or can function as more than just being "value vocabularies" in the Semantic Web.

The cases collected in this section demonstrate some innovative ideas that could be followed as relevant approaches to enhance the LOD KOS usage. Note that this section is not discussing semantic search and content discovery in a database or a website that is enabled by using KOS; here the cases are about the KOS themselves. They illustrate how LOD KOS can be *potentially* useful to researchers among the end-users, as found in the following situations: 1) using well-developed KOS products, high quality and relevant knowledge bases are now easily available for researchers; 2) name authorities could offer foundational structured data for network analyses; and 3) user-friendly displays of KOS provide visually enriched understanding.

**RS-1. Accessing and using KOS-based knowledge bases**

Researcher RS-1 needs to access and obtain information resources that could help answering sophisticated questions through a user-friendly workflow and tool. RS-1 has little knowledge of RDF or SPARQL. Fortunately, a countable number of innovative LOD KOS providers have provided user-friendly templates for querying their LOD KOS data. From the following examples, it is clear that researcher RS-1 can use these templates to obtain special graphs or datasets for very complicated questions.

The first example is from the Universal Protein Resource (UniProt[31]), a comprehensive resource for protein sequence and annotation data (see Figure 7). Organisms are classified in a hierarchical tree structure. The taxonomy database contains every node (taxon) of the tree. Top nodes are Archaea, Bacteria, Eukaryota, and Viruses. The UniProtKB taxonomy data is manually curated: next to manually verified organism names, a selection of external links, organism strains and viral host information are provided. Using the template, for example, in question #8, one can find all preferred gene name and disease annotation of all human UniProt entries that are known to be involved in a disease. This is much more complicated than the question #2, "Select all bacterial taxa, and their scientific name, from the UniProt taxonomy." In both cases, clicking on "show," will automatically load and make the query ready for use (see Figure 7).

---

[31] http://sparql.uniprot.org/

Figure 7. Query examples provided by UniProt (upper figure) and the SPARQL query for question #8, automatically "show"ed (lower figure).
Source: http://sparql.uniprot.org/.

One may argue that UniProt itself is a knowledge base, and the taxonomy is just used for organizing the information, raising the question as to whether a LOD KOS dataset itself could be considered as a knowledge base. The next case is the *Getty Thesaurus of Geographic Names* (TGN) available through Getty Vocabulary LOD service [32]. The application has turned the thesaurus into a knowledge base. For example, in combination with the geographic boundary data that are available in TGN, query #4.16 to #4.19 help to gather data such as places by, within, or outside a coordinate bounding box, and even with further criteria such as filtering by place type and obtaining geo or column charts (see Figure 8).

Figure 8. Templates of TGN-specific queries, provided by Getty Vocabularies LOD service. Source: http://vocab.getty.edu/queries#TGN-Specific_Queries

To demonstrate, the screenshot of Figure 9 is an action to obtain a dataset of "Places by Type Within Bounding Box". By choosing query #4.18 (left), the query template appears accordingly (lower right) and fills in the query box on top with a single click. The example provided by the template is to look for castles around the Netherlands (within 50.787185 3.389722 53.542265 7.169019). Now, it is at the hands of the researcher RS-1 to decide what "type" and what geographic boundary box he/she would like to check. For example, at first RS-1 replaced "castles" with "caves" and marked the geo coordinators around the ancient Silk Road, within 24.75083 28.95778 43.80722 108.92861; then RS-1 submitted the query (Figure 9 upper). The result was a dataset of over 200 caves spread in various countries (Figure 9 lower), all done within a few minutes (Zeng and Hu 2017). Each URI also brings the full data for each cave and other related information. The dataset is available for downloading with various formats.

Figure 9. Using the template provided by the LOD service, a query is submitted (upper figure), resulting a dataset (lower figure) for a specific place type (e.g., caves) in a geographic boundary. Source: http://vocab.getty.edu/queries#TGN-Specific_Queries

**RS-2. Name authorities offer foundational structured data for network analyses**

Researcher RS-2's attention was on the artists who played significant roles in history. Similar to the above, templates will help researcher RS-2 to use the *Union List of Artist Names* (ULAN) through Getty Vocabulary LOD service. The templates have also provided example queries for many complicated research questions which provide answers at impressive speeds. Checking each sample query in Figure 10, it reveals without a doubt that any answer to such a question would not be possible by simply searching or browsing on a website by an end-user. For example, now RS-2 can gather the datasets for all the Female Artists (#5.3), for Architects born in the 14th or 15th Century (#5.7), for Non-Italians who worked in Italy (#5.9), or for all kinds of data related to an artist's network, region, time period, cultural group. These are based on the established entries that have been carefully created and quality controlled by the KOS producers; hence the results have high quality.

Figure 10. Templates of ULAN-specific queries, provided by Getty Vocabularies LOD service.
Source: http://vocab.getty.edu/queries#ULAN-Specific_Queries

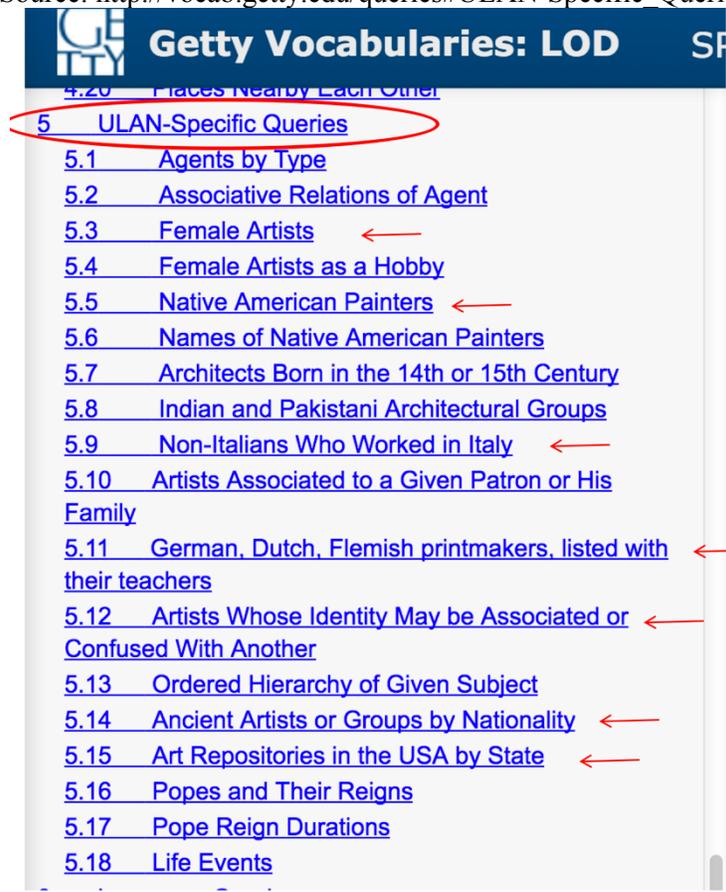

The following example is a query for finding associative relationships for *Wright, Frank Lloyd (American architect, 1867-1959)*, showing *relationship type, associated persons, each person's preferred name, preferred display biography*, and other *comments*. Again, the provided template #5.2 "Associative relations of Agent" made it possible for any end-user to just replace the URI of the aimed artist (e.g., ulan:500020307 for Frank Lloyd Wright), submit the query (Figure 11 upper), and get the results as datasets (Figure 11 lower) (Zeng 2017).

Figure 11. Using the template (upper figure) provided by the LOD service, a query is submitted to get the dataset for an artist *Wright, Frank Lloyd* and his associative relationships (lower figure). Source: http://vocab.getty.edu/queries#ULAN-Specific_Queries

The 37 related agents (Figure 11, lower figure) around this artist reveal the specific relationships. RS-2 or any user can further explore any of these related people, each named with a unique URI. If downloading the dataset (e.g., csv), one can also use other open tools (e.g., Google Fusion Tables, Gephi) to visualize the relationships with dynamic graphs.

We should also realize the importance of these URIs. Searching on the Web using such a URI, e.g., "ulan:500020307", the results will retrieve this artist's Wikipedia pages in all languages, the DBpedia entry, the links to the museums which host the artists' works (such as MoMA https://www.moma.org/artists/6459), and the libraries that have books about this artist (such as University of Wisconsin - Madison Libraries).

**RS-3. User-friendly displays of KOS provide visually enriched understanding**

To an end-user like researcher RS-3 who is not familiar with a KOS' structure and contents, a user-friendly display of KOS may provide visually enriched understanding. The *Cadastre and Land Administration Thesaurus* (CaLAThe[33]), is reported to have been derived mainly from the *ISO/DIS 19152 Land Administration Domain Model* and is related to existing thesauri, primarily the *GEMET thesaurus*, the *AGROVOC thesaurus*, and the *STW Thesaurus for Economics* (Volkan and Stubkjær 2015). The approach is similar to the case related to VP-3. The additional effort is that the service's graphical overviews render the main groups (Documentation, Land, Law, Party, and Activity) with thesaurus terms and relations. Individual concept searches also carry the results enriched with graphical views of the semantic relationships (see Figure 12).

Figure 12. The graphic overview of the group "Activity"
of the *Cadastre and Land Administration Thesaurus* (CaLAThe).
Source: http://cadastralvocabulary.org/

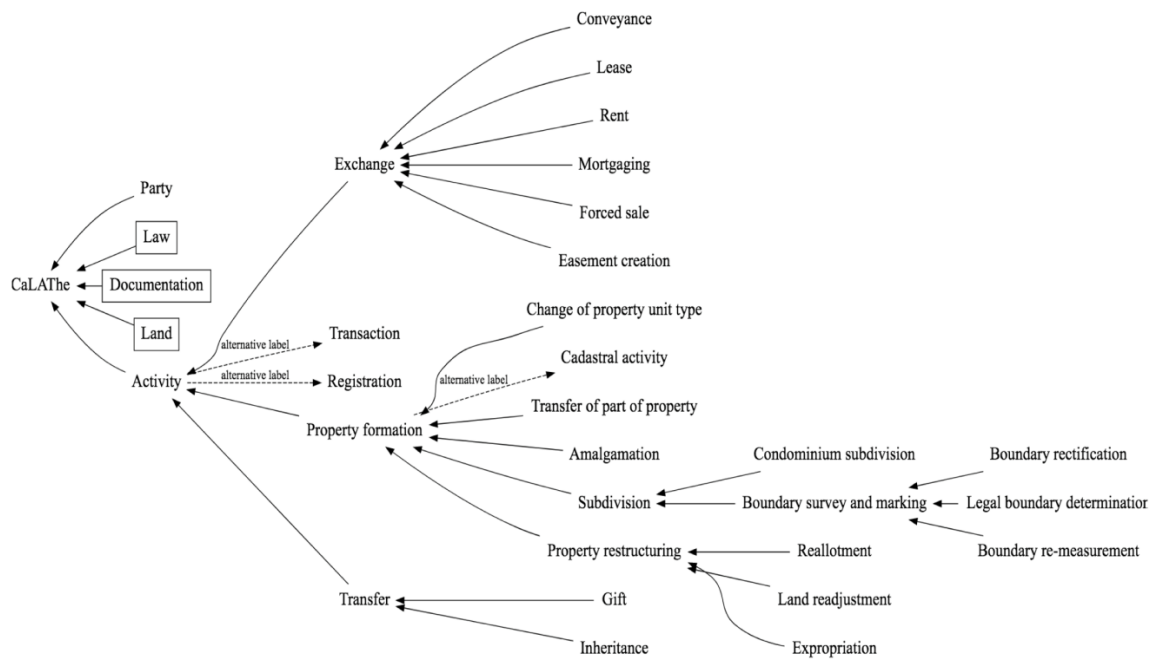

---

Tools like SKOS-play[34] are free applications to render and visualize thesauri, taxonomies or controlled lists expressed in SKOS. For the user who is not familiar with markup languages, the tool provides a way to convert Excel spreadsheets to SKOS files plus the visualization. Such features extend the benefits of SKOSified KOS publishing tools such as Skosmos[35] that would allow a vocabulary producer to test and verify a vocabulary during the conception phase; to exchange and communicate the vocabulary when validating it with domain experts; and to publish it when it is shared on the Web. With the added visual display function, end-users (not necessarily dataset or vocabulary producers) are able to have a visually enriched understanding of a KOS vocabulary's structure and contents.

**Summary and discussion (Researchers)**

The cases demonstrated in this sub-section highlight the great and endless potential of LOD KOS to be used by Researcher (RS) user group. The semantic rich structure and high quality controlled vocabulary now can be used in an innovative manner; further than the existing controlled vocabularies or standardized name authorities.

Additionally, the appropriate practices for the implementation, extension, access, and use of these standards in the final deliverables is critical to the real extended functionality of the KOS beyond being the controlled vocabularies or standardized name authorities. There is still a long way to go to make KOS be recognized as knowledge bases and semantic tools. It is important to realize the limitations of both typical web-based searching (simple, term-based) and browsing because these traditional methods are not taking the full advantages of *machine-processable* data that are much more powerful and useful than the previous *machine-readable* status.

# 5. Conclusion

This quote by Aristotle, "The whole is greater than the sum of its parts" reminds us how much better things are together than as separate pieces. It also applies to the principles of design. All the cases presented here, as the representatives of ideas and practices, demonstrate that although it is possible to use each available component of KOS independently, the real power lies in the skillful coordination of all. On the side of semantic technologies, the Semantic Web standards such as SKOS, OWL, RDFS, and SPARQL have paved the way for the conventional KOS to become LOD datasets. On the side of the information and knowledge professionals, there have been tremendous and continuous needs for KOS of all kinds, across domains, and worldwide. When the two sides embrace and when KOS join the mainstream in the 21st century, the opportunities for using the semantic-rich LOD KOS is much greater than ever before, due to the fact that LOD KOS data are *machine-understandable, -processable*, and -*actionable* (instead of just being *machine-readable*) in the Semantic Web, which connects *things* instead of *strings*.

In the effort to sort out the ideas and products related to LOD KOS (whether producing or using them) from disparate resources, this research first created personas as typical users of LOD KOS, to build a common understanding of the needs and goals various user groups want to achieve. The accumulated set of cases collected by this study is open-ended and the sources of this are unconventional, as explained in Section 3.1 "Method". The research aims at examining the functional changes that optimize the usage of LOD KOS from multiple dimensions, in order to share the practices and ideas among the related communities and users.

The findings indicate that the primary reason that LOD KOS vocabularies have become a fundamental component of the LOD building blocks is that they enable datasets to become 4- and

---

5-star Open Data. When trying to reach the benchmarks, every LOD Dataset Producer (DP) will realize their dependence on KOS which are their value vocabularies and the sources of URIs/IRIs to be used in data-transformation. The openly available, well-established, and constantly-maintained vocabularies are invaluable engines for the LOD datasets. The common issues and benchmarks summarized based on the study can be applied to any project that LOD dataset producers may encounter.

In the section for the Vocabulary Producer (VP) group, the major conceptual and structural methodologies used by the cases resemble some found in the history of KOS before the Semantic Web era. What makes them different is that the new approaches are empowered by the semantic technologies while the results comply with LOD principles. The data-driven, shared editing and publishing workflow also facilitates the capture of administrative, provenance, and use metadata for the whole KOS and its components. With more KOS being published in standardized, machine-understandable RDF format, institutions can improve and expand the KOS data that they already have with other outside sources. The most important achievement is the reusability of any of these new vocabularies.

The last section for researchers as end-users (RS) reveal great and endless potentials of LOD KOS. The semantic rich structure and high quality vocabulary now can be used integrated and innovative, on top of being the controlled vocabularies or standardized name authorities. LOD KOS datasets should be considered as knowledge bases, as the foundation of network analyses, and as the building blocks of a framework for research in the humanities and science. This might become the newest and most important function of KOS, although such cases are still rare. The authors of this paper believe that the barrier resides in communication about KOS through a delivering service rather than in the structure, format, or contents of a KOS.

The authors of the paper would like to call for more needed collaborations between the knowledge organization communities and the semantic technology communities. Meanwhile, researchers who are real end-users will be invaluable in such collaboration because their domain expertise, information needs, and information-seeking behaviors will lay out the questions that KOS knowledge bases can aim to answer, helping the growth of the KOS user communities with a variety of new objectives.

## Acknowledgements

We want to thank all reviewers for their positive and constructive comments which helped to improve this paper. In addition, we thank all our co-organizers of former NKOS workshops and all participants of NKOS-related events for their continuously input and feedback which motivated us to write this paper.
Supplementary materials (e.g. high resolution figures) of this paper are available under <https://github.com/PhilippMayr/supplementary-materials/tree/master/KOS-SW-MultidisciplinaryReview>.

## Appendix A. User persona document example: Vocabulary Producer (VP)

| Name | Vocabulary Producer | |
|------|---------------------|---|
| **Key** | VP | |
| **Sources** | *Original sources* | *Used for* |
| | LOV on Google+ https://plus.google.com/u/0/communities/108509791366293651606 | VP-1 |
| | Getty Vocab Google Group https://groups.google.com/forum/#!forum/gettyvocablod | VP-1 |
| | LODLAM challenges and sessions http://lodlam.net/ | VP-2 |
| | Research-based journal publications; conference and workshop presentations | VP-2, VP-3, VP-4 |
| | Theses and dissertations | VP-2, VP-3 |
| | GitHub entries such as OpenSKOS, NatLibFi/Skosmos, JSKOS | VP-4 |
| | Social media sources: tweets, blogs, Facebook groups | VP-2, VP-5 |
| | Informal interviews and local meetings | VP-1, VP-2 |
| | Mailing lists within a user group | VP-1 |
| **Tasks** | Vocabulary producers are involved in the development, maintenance, and enrichment of new and existing KOS in a wide range of scales (e.g., micro, satellite, unified, heterogeneous, extended, enriched, or other kinds).  The tasks usually include:<br>• Creating, developing;<br>• Maintaining, enriching, extending, translating;<br>• Integrating and unifying;<br>• Transforming (e.g., making an ontology from a thesaurus);<br>• Mapping with others;<br>• Sharing, reusing, contributing;<br>• Quality control and maintenance. | |
| **Content** | • Entries / instances -- with all property components required, including semantic and linguistic, format requirements, following standards and best practices;<br>• URIs – with namespace of any entry from any source;<br>• Rights and contributors;<br>• Provenance data;<br>• Updates info (new concepts, terms, relations, sources, etc.);<br>• Samples, previews, feedback, issues;<br>• Related images;<br>• Sources and URIs of the related real things;<br>• Alignments coded with appropriate degrees. | |
| **Interactions** | • Working platforms (spreadsheet, local database, open tool, etc.);<br>• Desktops /Mobile Applications;<br>• Websites (HTML, navigate-able);<br>• API-based services;<br>• SPARQL endpoints (with or without templates);<br>• Datasets. | |
| **Goals** | • Create and maintain high quality vocabularies;<br>• Follow the vocabulary principles of user-warrant, literary-warrant, organizational warrant;<br>• Follow international standards for KOS structure, components, and interoperability;<br>• Comply with Linked Data principles;<br>• Enrich, extend, and update contents constantly;<br>• Share, reuse, and contribute (both in and out) in vocabulary productions. | |